\documentclass{jpp}
\usepackage[utf8]{inputenc}
\usepackage{graphicx}
\usepackage{epstopdf, epsfig}
\usepackage{hyperref}
\usepackage{amssymb,amsmath}
\usepackage{bm}
\usepackage{verbatim}
\usepackage{tabularx}
\usepackage{color}

\usepackage{authblk}

\title{Formation of Transient High-$\beta$ Plasmas in a Magnetized, Weakly Collisional Regime}

\author{T. Byvank\aff{1},
	D. A. Endrizzi\aff{2},
	C. B. Forest\aff{2},
	S. J. Langendorf\aff{1},
	K. J. McCollam\aff{2},
	\and\ S. C. Hsu\aff{1} }
	
\affiliation{
	\aff{1}Physics Division, Los Alamos National Laboratory,
	Los Alamos, NM 87545, USA
	\aff{2}Wisconsin Plasma Physics Laboratory, Department of Physics, University of
	Wisconsin, Madison, WI 53706, USA}

\begin{document}
	
	\maketitle
	
\begin{abstract}
We present experimental data providing evidence for the formation of transient ($\sim 20$~$\mu$s) plasmas that are
simultaneously weakly magnetized (i.e., Hall magnetization parameter $\omega \tau > 1$) and dominated by thermal pressure (i.e.,
ratio of thermal-to-magnetic pressure $\beta > 1$). Particle collisional mean free paths are an appreciable fraction of the overall system size.
These plasmas are formed via the head-on merging of
two plasmas launched by magnetized coaxial guns.  The ratio $\lambda_{gun}=\mu_0 I_{gun}/\psi_{gun}$ of gun 
current $I_{gun}$ to applied magnetic flux $\psi_{gun}$ is an experimental knob for exploring the parameter space of
$\beta$ and $\omega \tau$. These experiments were conducted on the Big Red Ball at the Wisconsin Plasma Physics
Laboratory. The transient formation of such plasmas can potentially open up new regimes for the laboratory
study of weakly collisional, magnetized, high-$\beta$ plasma physics; processes relevant to astrophysical objects and phenomena; and 
novel magnetized plasma targets for magneto-inertial fusion.  
\end{abstract}

\section{Introduction}

Weakly collisional plasmas (particle mean free paths $\lesssim$ system size) with ratio of thermal-to-magnetic pressure $\beta>1$ and, simultaneously, Hall magnetization parameter $\omega \tau >1$
(i.e., gyro-frequency greater than collision frequency), 
represent a frontier regime of laboratory plasma physics research.
By contrast, magnetically confined plasmas typically have $\beta\ll 1$ and $\omega\tau
\gg 1$, whereas inertially confined and high-energy-density plasmas
typically have $\beta\gg 1$ and $\omega\tau\ll 1$.
If plasmas with both $\beta,\omega\tau >1$ can be formed successfully, an interesting next step will be to attempt to generate
small-scale, tangled magnetic field with connection length much longer than the characteristic scale size of the plasma.  This paper focuses on the first step of forming and
characterizing a transient plasma with both $\beta>1$ and $\omega\tau>1$ in a laboratory setting.

One motivation for this work is to help
establish a new laboratory platform to study
the fundamental physics of weakly collisional,
magnetized, high-$\beta$ plasmas \citep{kunz20} as a foundational
aspect of the scientific discipline of plasma physics. These types of plasmas, while ubiquitous in the universe, are poorly understood in terms
of their basic stability and transport properties \citep{chandran98,Schekochihin2008}, on
both macro- and micro-scales. A better predictive understanding of the behavior
of these types of
plasmas could potentially shed light on the dynamics and evolution of magnetohydrodynamic
(MHD) turbulence \citep{Schekochihin2007, Schekochihin2009} and magnetostatic turbulence \citep{Ryutov2002,Ryutov04}, which are fundamentally different compared to the drift-wave turbulence of low-$\beta$, magnetically confined plasmas; on
astrophysical systems such as accretion flows
around black holes \citep{balbus98}, the intracluster medium (ICM) within galaxy clusters \citep{schekochihin05,peterson06},
molecular clouds \citep{Ryutov04,federrath13mnras,padoan14},
the interstellar medium where large-scale structures form \citep{Scalo2004,McKee2007}, stellar/solar winds
\citep{bruno13lrsp}; and on outstanding cosmological questions such as the origin (magnetogenesis) and amplification (dynamo) of magnetic fields \citep{kulsrud08}.
In many of these systems, the magnetic pressure is significantly less than the combined plasma thermal and kinetic ram pressures. However, all of these components can play a non-negligible role in the plasma dynamics.

A second motivation for this work is to discover how to form a novel, magnetized, $\beta>1$ target plasma for magneto-inertial fusion (MIF), also known as magnetized target fusion (MTF) \citep{kirkpatrick95,Lindemuth2009}. MIF is a class of pulsed fusion approaches,
where a liner compresses a magnetized target plasma,
in which the magnetic field reduces thermal transport and enhances fusion-charged-product
(e.g., $\alpha$ particles) energy deposition within the plasma fuel.
Many MIF efforts over several decades have focused on the use of 
$\beta \leq 1$ plasmas, e.g., spheromaks \citep{Bellan2000} and field-reversed configurations (FRC) \citep{steinhauer2011}, as the target plasma for subsequent liner compression.  However,
these $\beta\leq 1$ plasmas suffer from MHD instabilities that have precluded the attainment of robust fusion conditions.
This has motivated the consideration of magnetized plasmas that can avoid
MHD instabilities while still benefitting from magnetic thermal insulation \citep{Ryutov2009,Hsu2019}, which could potentially
be enabled by the $\beta,\omega\tau>1$ regime. The lifetime
of a $\beta>1$ plasma target will be limited largely by hydrodynamic
expansion rather than MHD instability growth, and thus fast
liner compression is required, which has been demonstrated
successfully in Magnetized Liner Inertial
Fusion (MagLIF) \citep{Slutz2010,gomez14} and is the aim of plasma-jet-driven MIF (PJMIF) \citep{thio99,Hsu2012,Thio2019}.  The present work is a first step
toward determining the viability of forming potential magnetized target plasmas
with $\beta,\omega\tau > 1$ that may be suitable for subsequent, fast liner compression.

In this work, performing experiments on the Big Red Ball (BRB) \citep{Forest2015} at the Wisconsin Plasma Physics Laboratory (WiPPL), we launch and merge two $\beta \sim 1$ plasmas
to transiently create $\beta,\omega\tau>1$ conditions for $\sim 20$~$\mu$s, and experimentally measure the plasma parameters in both the individual and merged plasmas.
Compared with individual plasmas, the head-on collisions (1)~increase the duration for which the desired plasma state exists at a particular location and (2)~increase the magnitudes of the density and radial and toroidal magnetic field components, which widens the parameter space compared to that achievable with individual plasmas. The transient nature is unavoidable for $\beta > 1$ plasmas unless they are wall confined, which would bring in complications including plasma-wall interactions and impurities. 
Our plasma-formation approach is analogous to many prior studies that
generated $\beta < 1$ spheromaks using 
magnetized coaxial guns \citep{Bellan2000} and the merging/collision of two $\beta < 1$ compact toroids, e.g., spheromaks \citep{Bellan2000} or field reversed configurations (FRC) \citep{steinhauer2011}, for magnetic-reconnection studies \citep{Yamada1990, Ono1999, Cothran2003} and fusion concept exploration \citep{Slough2011, Guo2011}.
To access various portions of the $\beta$, $\omega\tau$ parameter space, we tune the
experimentally adjustable parameter $\lambda_{gun} = \mu_0 I_{gun}/\psi_{gun}$, where $I_{gun}$ is the peak gun electrical current and $\psi_{gun}$ is the applied vacuum poloidal magnetic flux
linking the gun electrodes \citep{Yee2000,Hsu2005}. Engineering improvements to our plasma injectors are presently underway to improve reliability and reproducibility of plasma formation.

The paper is organized as follows.  Section~\ref{sec:background} provides
background information on the scaling of $\beta$ and $\omega \tau$ and the plasma-formation
process. Section~\ref{sec:setup} describes the experimental setup and diagnostics. Section~\ref{sec:results} presents experimental results for individual and merged plasmas. Section~\ref{sec:future} discusses future work. Section~\ref{sec:concl} provides conclusions.

\section{Background \label{sec:background}}

In this section, we provide background information that further motivates
this work and helps with understanding the experimental setup, methods and the
results presented later in the paper.

\subsection{Plasma parameters}

In this subsection, we describe the plasma parameters needed to achieve
$\beta,\omega\tau>1$ simultaneously.  The
definition of $\beta$ is
\begin{equation}
	\beta = \frac{P_{th}}{P_{mag}} = \frac{2 \mu_{0} \sum_{j} n_{j} k_B T_{j} }{B^{2}} \approx (4.0 \times 10^{-11}) \frac{\sum_{j} n_{j}[\mathrm{cm}^{-3}] T_{j}[\mathrm{eV}] }{(B[\mathrm{G}])^2}  \propto \frac{n_i (ZT_{e} + T_{i}) }{B^{2}},
	\label{eqn:beta}
\end{equation}
where $P_{th}$ is the plasma thermal pressure, $P_{mag}$ the magnetic pressure, $\mu_{0}$ the vacuum permeability, $n_{j}$ the ion or electron density ($j=i,e$), $k_{B}$ the Boltzmann constant, $T_{j}$ the ion or electron temperature, $B$ the magnetic field strength, and $Z$ the mean ion charge state.
A thermal-pressure-dominated plasma has $\beta > 1$.
The definitions of the ion and electron Hall magnetization parameters, $\omega_i \tau_i$ and $\omega_e \tau_e$, respectively, are
\begin{equation}
	\omega_{i}\tau_{i} = \frac{\omega_{i}}{\nu_{i}} \approx (2.0 \times 10^{11}) \frac{B[\mathrm{G}] (T_{i}[\mathrm{eV}])^{3/2}}{Z^{3} \mu^{1/2} n_{i}[\mathrm{cm}^{-3}] \ln{\Lambda}} \propto \frac{B T_{i}^{3/2}}{Z^3\mu^{1/2}n_i}
	\label{eqn:mag_i}
\end{equation}
and
\begin{equation}
	\omega_{e}\tau_{e} =\frac{\omega_{e}}{\nu_e} \approx (6.0 \times 10^{12}) \frac{B[\mathrm{G}] (T_{e}[\mathrm{eV}])^{3/2}}{n_{e}[\mathrm{cm}^{-3}] \ln{\Lambda}} \propto \frac{B T_{e}^{3/2}}{n_e}
	\label{eqn:mag_e},
\end{equation}
where $\omega_{i,e}$ are the ion and electron gyrofrequencies, $\tau_{i,e}$ the
ion and electron collision times, $\nu_{i,e}$ the ion and electron collision rates,
$\mu$ the atomic mass number, and $\ln{\Lambda} \approx 10$ the Coulomb logarithm. The condition $\omega_{i}\tau_{i}~>~1$ is usually more stringent than $\omega_{e}\tau_{e}~>~1$.

\begin{figure}
	\includegraphics[width=\textwidth,keepaspectratio]{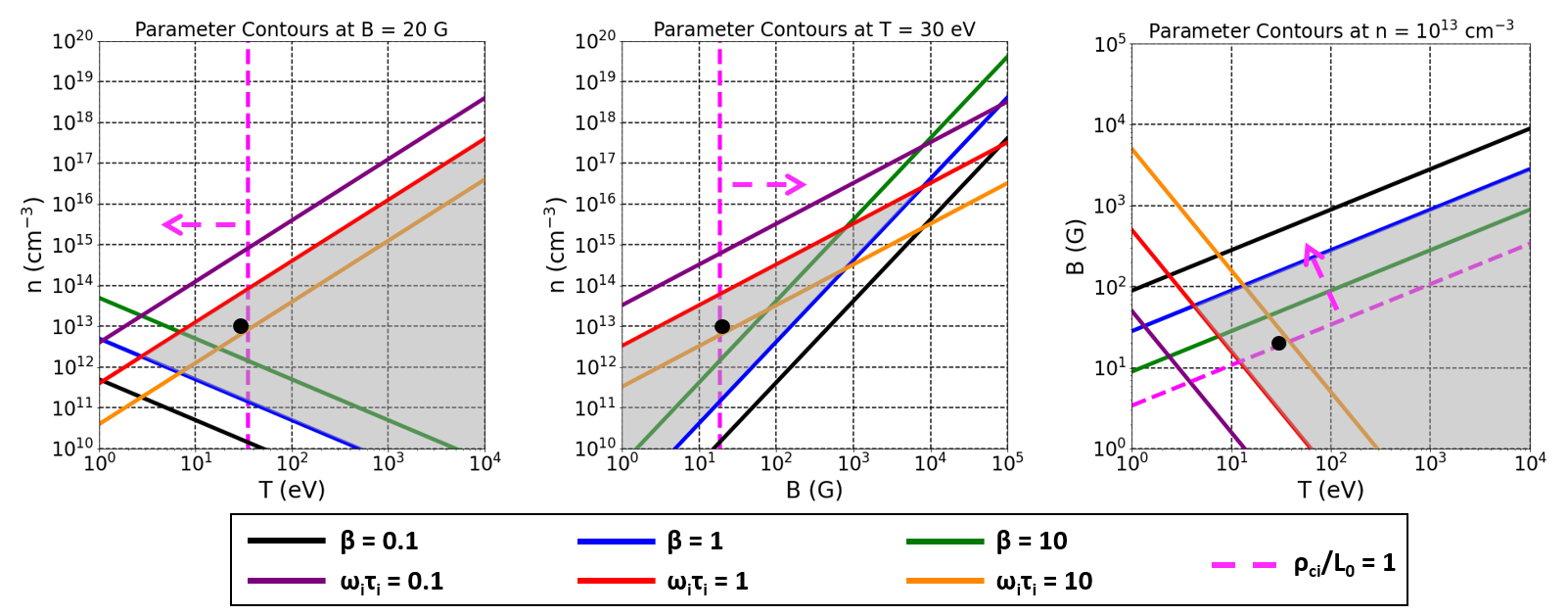}
	\caption{
		Contours of $\beta$, $\omega_{i}\tau_{i}$, and $\rho_{ci}/L_{0}$ in the (left) $n$-$T$ plane at constant $B = 20$~G, (middle) $n$-$B$ plane at constant $T=T_{i} = T_{e} = 30$~eV, and (right) $B$-$T$ plane at constant $n=n_{i} = n_{e} = 10^{13}$~cm$^{-3}$. The shaded regions correspond to the desired regime of $\beta, \omega\tau >1$. The dashed arrows show the regime for which $ \rho_{ci}/L_{0} < 1$, with $L_{0} = 30$~cm.} The dots correspond to the approximate values obtained in the present research.
	\label{fig:contour_space}
\end{figure}

Figure~\ref{fig:contour_space} illustrates contours of $\beta$,$\omega_{i}\tau_{i}=0.1, 1$, and 10 as a function of $n$, $T$, and $B$, assuming that $n=n_i=n_e$ (and
therefore $Z=1$), $T=T_{i} = T_{e}$, and $\mu =$ 1. The shaded regions in the plots denote the parameter spaces where $\beta, \omega_i\tau_i >1$.  Because
$n$ is in the numerator of $\beta$ and denominator of $\omega\tau$, and vice versa
 for $B$, in general there are only limited ranges 
and combinations of $n$, $T$, and $B$ for which
$\beta$ and $\omega\tau$ can be simultaneously greater than unity. In 
general, higher $T$ helps because $T$ is in the numerator for both $\beta$
and $\omega\tau$. Additionally in Fig.~\ref{fig:contour_space}, we plot the ratio of the ion cyclotron radius $\rho_{ci} \propto T_{i}^{1/2}/B$ to the characteristic plasma radius $L_{0} = R = 30$~cm, here. The dashed arrow shows the region in parameter space for which the characteristic plasma radius is larger than the ion cyclotron radius, $ \rho_{ci}/L_{0} < 1$. The dots in the plots denote approximately the values obtained experimentally in this research.

\subsection{Plasma formation}

In this subsection, we describe the plasma-formation method and insights derived from previous work that we exploit to achieve the objectives of the present research.
It has long been known that compact-toroid (CT) plasmas, i.e., 
spheromaks and FRCs, are
formed in the $\beta \leq 1$, $\omega\tau>1$ regime, where the thermal pressure is confined by an equal or larger magnetic pressure. 

\begin{figure}
	\includegraphics[width=\textwidth,keepaspectratio]{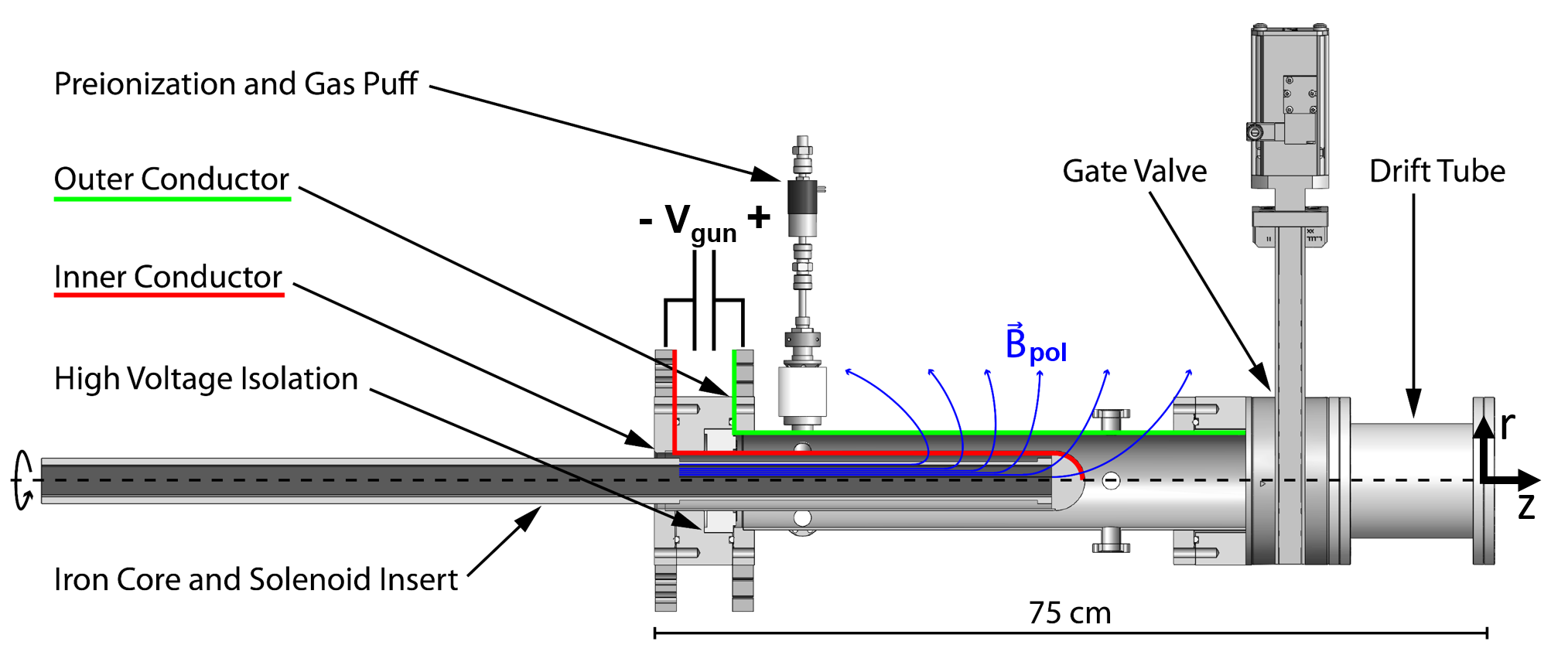}
	\caption{
		Diagram of a plasma gun/injector used in this research, showing the coordinate system, inner (cathode, red) and outer (anode, green) electrodes, and solenoid and iron core that generate the bias poloidal magnetic flux (blue). Gas is injected between the electrodes and pre-ionized using a separate ``washer gun'' system \citep{Fiksel1996}.  Finally, a discharge with gun
		voltage $V_{gun}$ between the electrodes accelerates the plasma out of the injector in the
		$z$ direction.
	}
	\label{fig:injector}
\end{figure}

Spheromak and spheromak-like plasmas can be created using magnetized
plasma guns with coaxial, cylindrical electrodes in the presence of an
applied ``bias'' poloidal magnetic flux $\psi_{gun} = \int B_{pol} \cdot dA$ linking the two electrodes (see Fig.~\ref{fig:injector}). Electrical current $I_{gun}$ in
the electrodes ($z$ direction) and in the plasma ($r$ direction) that forms between
the electrodes generates a toroidal magnetic field ($B_{tor}$, in the $\phi$ direction), and the associated magnetic pressure accelerates the plasma out of the electrodes ($z$ direction). The moving plasma advects the poloidal bias magnetic field ($B_{pol}$, in the $r$-$z$ plane). Depending on the values of $I_{gun}$ and $\psi_{gun}$,
the magnetic field lines associated with the plasma can reconnect and detach from the electrodes as the plasma propagates out of the gun. The resulting plasma has both (1)~$B_{pol}$ associated with toroidal currents and (2)~$B_{tor}$ associated with poloidal currents. The relative values of magnetic field, density, and temperature in the plasma, and their evolution, determine the values of $\beta$ and $\omega\tau$. For $\beta > 1$, the magnetic field is not large enough to create an equilibrium with the plasma thermal pressure, so $\beta>1$ plasmas will exist only transiently.

The $\lambda_{gun}$ parameter,
\begin{equation}
	\lambda_{gun} = \frac{\mu_{0} I_{gun}}{\psi_{gun}} \propto \frac{B_{tor}}{\int B_{pol} \cdot dA},
	\label{eqn:lambda_gun}
\end{equation}
determines whether a spheromak forms \citep{Bellan2000,Yee2000,Hsu2005}.
A spheromak ($\beta < 1$, $\omega\tau>1$) will be formed under the condition \citep{Bellan2000}
\begin{equation}
	\lambda_{gun} > \lambda_{crit} \approx 3.83/r_{gun},
	\label{eqn:lambda_crit}
\end{equation}
where 3.83 is the first root of the Bessel function of the first kind $J_{1}(x)$, and $r_{gun}$ is the characteristic size (e.g.,
inner radius of the outer gun electrode). On the other hand, a plasma jet with rapidly
decaying magnetic flux is formed when $\psi_{gun} = 0$ and $\lambda_{gun}=\infty$ \citep{Hsu2012PoP,Merritt2014}, resulting
in a plasma with $\beta\gg 1$ and $\omega\tau\ll 1$ after a few
resistive decay times of the magnetic flux (on the order
of several $\mu$s). Given these two bracketing conditions,
it is intuitive to expect that a plasma with both $\beta,\omega\tau > 1$ may occur for some $\lambda_{gun}$ satisfying $\lambda_{crit} < \lambda_{gun} < \infty$.

In the present work, we launch and collide (head-on) two plasmas formed
by coaxial guns. We orient the applied bias magnetic fields such that the plasmas have opposite magnetic helicities $H$, where $H= \int \mathbf{A} \cdot \mathbf{B}\,\mathrm{d}V$ and $\mathbf{A}$ is the magnetic vector potential with $\mathbf{B}=\nabla \times \mathbf{A}$. In the $\lambda_{gun}\gtrsim \lambda_{crit}$ regime, the helicity of merging CTs determines the magnetic topology of the resultant plasma, i.e., the merging of two co-helicity or counter-helicity spheromaks creates a
$\beta \ll 1$ spheromak or a $\beta\lesssim 1$ FRC, respectively \citep{Yamada1990, Ono1999}. However, this precludes the achievement of
the $\beta,\omega\tau>1$ regime that we seek. The goal of this
research is to explore and characterize the head-on merging of two coaxial-gun-formed plasmas in the range $\lambda_{crit}<\lambda_{gun}<\infty$ in order to
form a transient plasma with $\beta,\omega\tau>1$.

\subsection{Plasma propagation and merging}

As the plasmas propagate into the vacuum chamber from the magnetized coaxial guns, they expand into an applied background magnetic field that is oriented along the propagation direction ($z$ axis). The
applied background field mitigates the amount of expansion. 
To estimate the range of plasma parameters during plasma propagation, we use a combination of measurements at the chamber center and near the gun, along with basic scaling relationships. This expected range of plasma parameters dictates the range of $\beta$, $\omega\tau$, Mach number $M$, Alfv\'en Mach number $M_A$, and mean free paths, as well as the nature of the plasma-merging dynamics. Further details are provided in Sec.~\ref{sec:results_prop}.

\section{Experimental setup \label{sec:setup}}

In this work, we launch two $\beta > 1$ plasmas head-on within a vacuum chamber and measure the resulting plasma densities, temperatures, and magnetic fields.
The experiments are conducted on the Big Red Ball, which is part of the
Wisconsin Plasma Physics Laboratory (WiPPL), a frontier-plasma-science user facility.

\subsection{Experimental chamber: the Big Red Ball}

\begin{figure}
	\includegraphics[width=\textwidth,keepaspectratio]{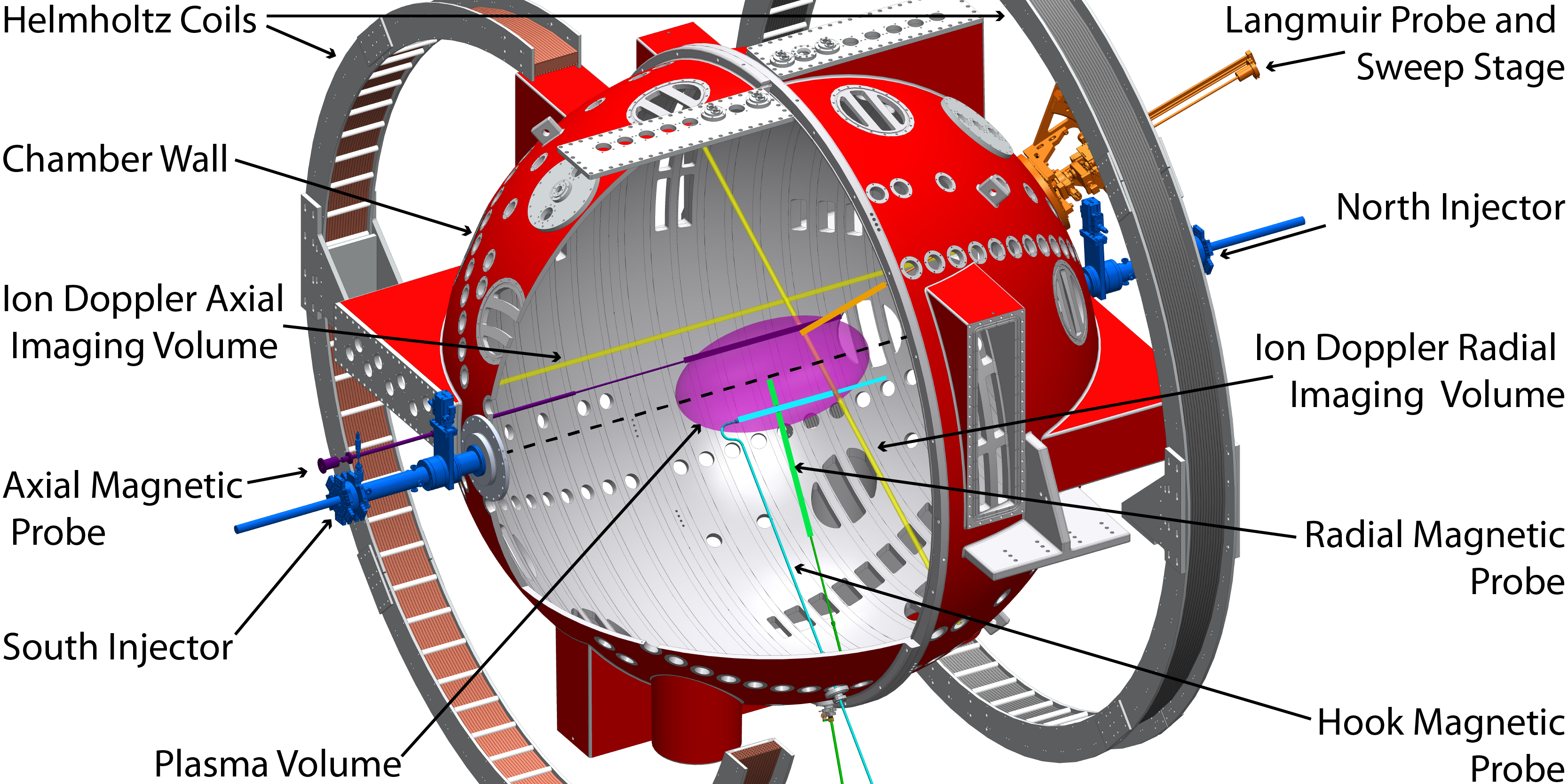}
	\caption{
		Diagram of the experimental setup, showing relative locations of the plasma injectors, diagnostics, approximate plasma volume in the 3-m-diameter BRB chamber, and external Helmholtz coils.
	}
	\label{fig:BRB}
\end{figure}

Figure~\ref{fig:BRB} shows a diagram of the Big Red Ball (BRB) experimental chamber at the WiPPL \citep{Cooper2014,Forest2015}. The multi-cusp magnetic confinement of the BRB contributes $<1$~G to the plasma-merging region. Two plasma injectors are positioned on opposite poles (labeled South and North) of the 3-m-diameter vacuum chamber at a base pressure of $\sim4$~$\mu$Torr. A 3-m-diameter Helmholtz coil set outside of the chamber provides a 50-G (DC) magnetic field pointing from south to north. This background magnetic field serves to prevent the $\beta>1$ plasmas from expanding too much and reducing in density before colliding at the chamber center. The magnetic field from the Helmholtz coil alters the magnetization of the iron core and thus the bias flux within the coaxial injectors, which is taken into account in the reported $\lambda_{gun}$ values. 

We define the chamber coordinates in the poloidal plane by $(r, z)$, where the south pole is
$(r = 0~\mathrm{cm}, z = -150~\mathrm{cm})$, and the north pole is $(r = 0~\mathrm{cm}, z = +150~\mathrm{cm})$. Diagnostics are placed at different toroidal angles $\phi$, and toroidal symmetry of the plasma structure is assumed for estimating plasma parameters. Experimental results suggest that this assumption may not always be valid. However, toroidal symmetry is not a requirement for accessing the parameter regimes for the physics studies of interest, nor for the studies themselves.

\subsection{Plasma injector}

Figure~\ref{fig:injector} illustrates the coaxial plasma injectors \citep{Matsumoto2016, Matsumoto2016a, EDO2018} that 
we use to create our plasmas. The inner radius of the outer electrode (anode) is $r_{gun} = 4.15$~cm, leading to $\lambda_{crit} \approx 92~$m$^{-1}$ for spheromak formation. Thus, we desire $\lambda_{gun}>\lambda_{crit}$, by reducing $\psi_{gun}$, in order to create $\beta>1,\omega\tau>1$ magnetized plasmas.

The plasmas are created by (1)~establishing a poloidal bias flux $\psi_{gun}$ between
the coaxial electrodes, (2)~injecting neutral gas between the electrodes, (3)~pre-ionizing the gas to create a plasma, (4)~accelerating the plasma by discharging current through the electrodes.
To establish $\psi_{gun}$, an iron core surrounded by a copper winding is inserted into the inner coaxial electrode (cathode). A current $<3$~A (DC) through the 4.1~cm diameter winding supplies $\psi_{gun}<0.4$~mWb within the $\sim$30~cm length region of plasma formation and acceleration in the coaxial injector. To vary $\lambda_{gun}$, we vary $\psi_{gun}$ and fix $I_{gun}$.
Gas is injected by valves into the coaxial gun 3~ms before the trigger of the 
main electrode discharge. 

In the present work, we use H and He gases for the South and North injectors, respectively.  Helium is used for ion Doppler spectroscopy measurements. We only use H in the South injector, for which we observe better performance with H than with He (not yet understood). The South injector is on loan from TAE Technologies, Inc., and the North injector is modified, designed, and built at WiPPL based on the TAE injector.
The injected gas diffuses through a ``washer gun'' \citep{Fiksel1996}, in which an applied pre-ionization voltage $<1$~kV breaks down the gas, and a 1-kA current sustains 100~kW of heating power. The washer guns use a $<1$-kG magnetic field produced with a small solenoid to assist in plasma breakdown. This field is roughly aligned with the poloidal magnetic field produced by the iron-core solenoid.

The main current $I_{gun}$ that accelerates the plasma out of the coaxial gun by the $\mathbf{J}\times \mathbf{B}$ force has a peak of $\approx 130$~kA and
a risetime of 5~$\mu$s. The current is crowbarred to prevent ringing. For each injector, the main discharge circuit has a capacitance of 70~$\mu$F operating at voltages of 10~kV, thereby storing 3.5~kJ. Accounting for losses in the transmission line, the energy delivered to the injector is $\sim 0.5$~kJ, estimated from time-resolved current and voltage measurements. After the plasma is accelerated along the length of the electrodes, it travels $\sim$35~cm through the injector and drift tube before it enters the experimental chamber.

\subsection{Diagnostics}

\subsubsection{Visible fast-framing camera imaging}

We image the individual plasmas and plasma collisions using a fast-framing Phantom v710 camera with 1-$\mu$s exposure times and 3-$\mu$s inter-frame time. The camera is positioned at an approximately 45$^\circ$ angle relative to the axial plasma-propagation direction. Narrow bandpass filters selectively admit H$_\alpha$, H$_\beta$, He-\textsc{i}, and He-\textsc{ii} plasma line emission. The imaging helps verify magnetic-signal timings and with experimental troubleshooting.

\subsubsection{Langmuir probe}

Electron densities and electron temperatures are measured using a multi-tip Langmuir probe. 
The 16 tips each have independent bias voltages to sample the current-voltage ($I$-$V$) traces with $0.2$-$\mu$s sampling period. The position of the Langmuir probe within the chamber can be adjusted, but for the data presented in this work, the probe was fixed at the chamber coordinates $(r~=~25~\mathrm{cm}, z~=~10~\mathrm{cm})$. The $n_{e}$ is given by
\citep{Cherrington1982}
\begin{equation}
n_{e} = \frac{I_{sat}}{0.61 A_{p} e} \Big(\frac{\mu m_{p}}{k_{B} T_{e}} \Big)^{1/2}
\label{eqn:Langmuir_ne}
\end{equation}
where $I_{sat}$ is the ion saturation current, $A_{p}$ is the probe area, $e$ is the elementary charge, and $m_{p}$ is the proton mass. We use $\mu=1$, 4, and 2.5 for H, He, and merged H-He plasmas, respectively.
The $T_e$ is inferred by fitting
\begin{equation}
I = I_{sat} \left\{ \exp{\left[\frac{e(V - V_{f})}{k_{B} T_{e}}\right]} - 1 \right\}
\label{eqn:Langmuir_Te}
\end{equation}
to the exponential part of the $I$-$V$ curve.

The discrete nature of sampling with a multi-tip probe introduces slightly more error than a typical swept Langmuir probe.  
All error bars presented here represent statistical uncertainties introduced via measurement and propagated in the analysis, but do not represent the typically much larger errors from Langmuir probe theory \citep{hutchinson_2002}. Looking at the quiescent period after the plasmas have collided, we estimate the standard deviations in the measured $I$ and $V$ to be 0.2~V and 1~mA respectively. Using these values, we take the $I$-$V$ data and generate a $N = 10^4$ size population of mock data with standard deviations around those values. By repeating these calculations at different times and for different shots, we find a 4\% uncertainty in $T_e$ and a 5\% uncertainty in $n_e$, which are the combination of the (non-systematic) uncertainties in measurement and the fitting routine. These represent lower bounds in the error, as they only include noise and not theoretical or systematic error.

\subsubsection{Magnetic probe arrays}

The magnetic fields and velocities are measured using an array of Bdot probes (``Hook Magnetic Probe'' in Fig.~\ref{fig:BRB}),
where the probe loop voltage $V \propto \partial B / \partial t $. The Bdot probe array consists of 11 locations equally spaced between $(r = 16~\mathrm{cm}, z = -20~\mathrm{cm})$ and $(r = 16~\mathrm{cm}, z = 40~\mathrm{cm})$ for the data presented in this work. At each array location there are 6 loops--3 orthogonal pairs of oppositely wound loops--to measure the three spatial components of the magnetic field and minimize common mode noise. Data are recorded at a $0.1$-$\mu$s sampling rate and numerically integrated to provide
values of the magnetic field. The Bdot-coil areas are calibrated to within a few percent.
There is a systematic uncertainty of the entire probe array of $\sim 1$~cm.

There are two other Bdot-probe arrays in the chamber (``Axial Magnetic Probe'' and ``Radial Magnetic Probe'' in Fig.~\ref{fig:BRB}), which provide magnetic signals at different radii and toroidal positions. These arrays help 
determine the radial extent of the plasma ($\sim$30-cm radius) as a function of toroidal angle. We also observe $\sim 30$-km/s plasma radial expansion, which affects the ion Doppler spectroscopy data.

\subsubsection{Ion Doppler spectroscopy}

Ion temperatures are measured using ion Doppler spectroscopy. We observe Doppler broadening of the 468.6-nm He-\textsc{ii} line along a radial viewing chord at $z~=~10$~cm. 
Light is collected using a 2.54-cm collimator, a 10-nm-bandwidth monochromater, 3-$\mu$m fiber bundles, and fed into a Czerny-Turner type spectrometer, as described in \cite{DenHartog1994}. We obtain measurements at a sampling rate of $\sim 1$~$\mu$s. The measured broadening is a convolution of broadening from the following sources: the ion temperature, the spectrometer instrumental broadening, and the plasma radial-expansion velocity. The result is $\sigma^{2}_{measured} = \sigma^{2}_{Ti} + \sigma^{2}_{instr} + \sigma^{2}_{exp}$, where $\sigma$ is the Gaussian broadening, i.e., full-width at half-maximum (FWHM). The instrumental broadening at $T_i=30$~eV is equivalent to $\sigma_{instr}=$1.5~eV,
but the relative contribution increases for smaller $T_i$. Broadening from the $\sim 30$-km/s plasma radial expansion is equivalent to $\sigma_{exp} \sim$ 9~eV. As $T_e$ cools below $\sim 10$~eV, the emission drops rapidly, and $T_i$ cannot be measured.

\section{Experimental results \label{sec:results}}

\subsection{Estimated plasma parameters \label{sec:results_prop}}

The nature
of head-on merging of two plasmas, e.g., whether they interpenetrate or form
sonic or MHD shocks, depends on the plasma parameters just prior to
merging. As the plasmas propagate with axial speed $v \sim 70$~km/s into the chamber from the magnetized coaxial guns, they expand into a 50-G applied background magnetic field that is oriented along the propagation direction ($z$ axis).
The plasmas expand radially (with instantaneous radius $R$) and axially (with
instantaneous length $L$) each by a factor of $\sim 8$ from the initial size near the gun to the final size near chamber center, with volume $V \propto L R^{2}$.
Near chamber center, we measure the plasma $R$, $L$, $n_e$, $T_e$, $T_i$, $\mathbf{B}$,
and $v$.  In the drift tube near the gun, we measure $B$ 
during separate experiments, and we infer the other quantities by making assumptions
about the expansion scalings (i.e., conservation of particles and resistive diffusion of magnetic flux), and verify the consistency of the inferred parameters near the gun with that of prior work using similar plasma injectors \citep{Matsumoto2016, Matsumoto2016a, EDO2018}.

Making the assumptions of conservation of particle and exponentially decaying magnetic flux while the expanding plasma propagates from the gun to chamber center, we have the
following relations: $n \propto L^{-1}R^{-2}$; $B_{r}, B_\phi \propto L^{-1}R^{-1}e^{-Ct}$; and $B_{z} \propto R^{-2}e^{-Ct}$ with propagation time $t$ and a constant $C$. 
While it is tempting to assume adiabatic expansion ($PV^{\gamma}$ = constant, with $\gamma = 5/3$) to infer the dependence of $T$ on plasma volume $V$,
the strict adiabatic-cooling rate $T \propto L^{-2/3}r^{-4/3}$ would lead to unrealistic $>$~keV temperatures near the gun when extrapolated from measured temperatures near chamber center.
Instead, as a lower bound, we take 
$T_{i} \approx$ constant. At higher densities near the gun, $T_e$ and
$T_i$ equilibrate over $\sim 1$-$\mu$s timescales, but at the lower densities near
chamber center, the ion--electron equilibration time is longer than the $\sim 20$-$\mu$s propagation time. Thus, $T_e$ decreases from the initial $T_{e} \approx T_{i}$ near the gun to the measured $T_{e} < T_{i}$ near chamber center,
due to radiation and $P\mathrm{d}V$ work on the background magnetic
field. Concurrently, resistive dissipation of the magnetic field causes Ohmic heating of the electrons. To put bounds on parameters rather than to be predictive of the precise plasma temperature between the gun and chamber center, we simply model a linear decrease in $T_{e}$ during propagation.

\begin{table}
	\caption{Measured and estimated values of plasma parameters (for single plasmas) for the experiments
		reported in this paper, assuming $Z=1$, $\mu=1$, $\gamma=5/3$, and $\ln{\Lambda}=10$.}
	\label{tab:param}
	\begin{tabular}{lcc} \hline\hline
		Parameter & Near Gun & Near Chamber Center \\ \hline
		Propagation distance (cm) & 0 & 150 \\
		Time $t$ ($\mu$s) & 0 & 20 \\
		Plasma radius $R$ (cm) & 4 (est.) & 30 (meas.) \\
		Plasma length $L$ (cm) & 10 (est.) & 80 (meas.) \\
		Electron density $n_e$ (cm$^{-3}$) & $5 \times 10^{15}$ (est.) & $10^{13}$ (meas.) \\ 
		Electron temperature $T_e$ (eV) & 30 (est.) & 15 (meas.) \\
		Ion temperature $T_i$ (eV) & 30 (est.) & 30 (meas.) \\
		Magnetic field $B$ (G) & 2000 & 20 (meas.) \\
		Thermal electron collision time $\tau_e$ & 1~ns & 0.2~$\mu$s \\
		Thermal ion collision time $\tau_i$ & 68~ns & 34~$\mu$s\\
		$\beta$ & 3 & 45 \\
		$\omega_{i}\tau_{i}$ & 1 & 7 \\
		$\omega_{e}\tau_{e}$ & 40 & 70 \\
		Axial plasma speed $v$ (km/s) & 70 (est.) & 70 (meas.) \\
		Sound speed $C_s$ (km/s) & 70 & 50 \\
		Alfv\'en speed $V_A$ (km/s) & 60 & 14 \\
		Mach number $M$ & 1 & 1.4 \\
		Alfv\'en Mach number $M_A$ & 1.2 & 5 \\	
		Electron thermal mean free path $\lambda_{\rm mfp,e}$ (cm) & 0.3 & 30 \\
		Ion thermal mean free path $\lambda_{\rm mfp,i}$ (cm) & 0.4 & 180 \\
		Ion--ion interpenetration length for merging plasmas $L_{ii,s}$ (cm) & (N/A) & 57 \\	
		Ion gyroradius $\rho_i$ (cm) & 0.3 & 30 \\
		Ion inertial length $c/\omega_{pi}$ (cm) & 0.3 & 7 \\
		\hline
	\end{tabular}
\end{table}

\begin{figure}
	\includegraphics[width=\textwidth,keepaspectratio]{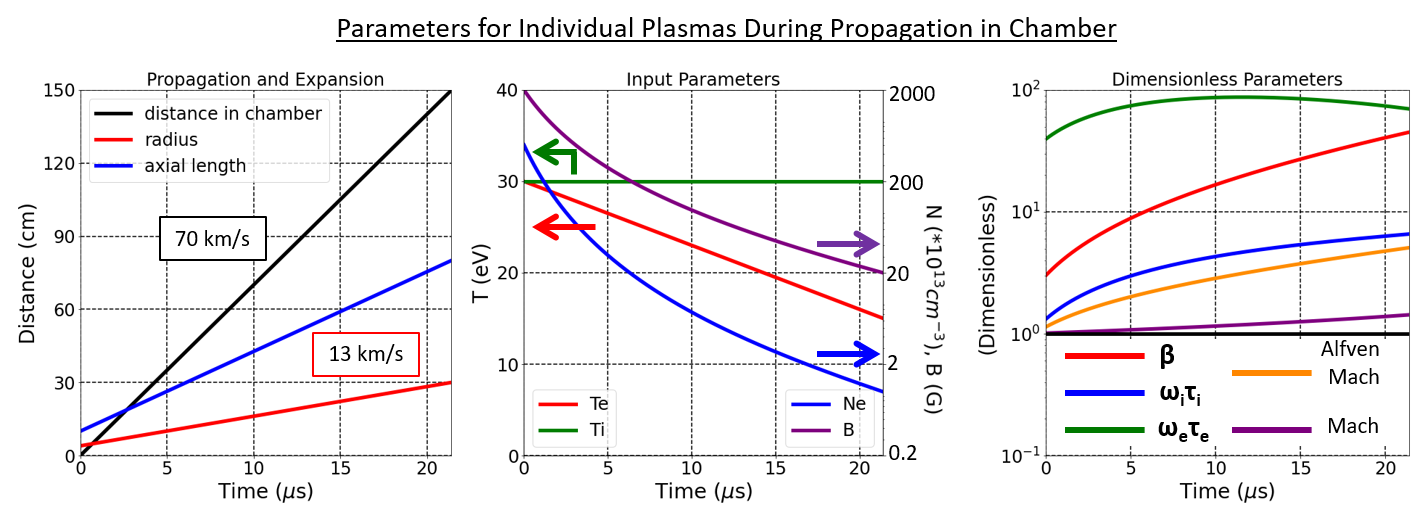}
	\caption{
		Anticipated plasma parameters vs.\ time for individual plasmas as they propagate to the center of the chamber. Values are based on interpolating the data in Table~\ref{tab:param} and serve to place bounds on parameters rather than to precisely predict behavior during propagation.
	}
	\label{fig:dist_plots}
\end{figure}

Based on the relations and information discussed above, we provide estimates
of parameters near the gun and near chamber center in Table~\ref{tab:param},
which suggests that
thermal pressure is greater than magnetic pressure (i.e., $\beta>1$) both
near the gun and near chamber center. While the radial expansion speed measured by the B-dot probe array near the chamber center is $\sim 30$~km/s, the average expansion speed calculated during propagation through the chamber (from Fig.~\ref{fig:dist_plots}) is $\sim 13$~km/s. Both observations of the radial expansion speed are less than the ion sound speed $C_s$ or Alfv\'en speed $V_A$ near the gun (in Table~\ref{tab:param}). Near chamber center, the measured expansion speed is still subsonic. During head-on merging, the plasmas are super-Alfv\'enic and slightly supersonic, and thus it may be
expected that MHD shocks could form.  However, during initial head-on merging
of two plasmas, the classical ion--ion Coulomb interpenetration length between the
merging plasmas is an appreciable fraction of $L$, and thus it may be expected that any shock formation will be delayed if
it forms at all, as observed in prior work with merging supersonic plasmas \citep{moser15pop,langendorf19pop}.
By interpolating the data in Table~\ref{tab:param}, we show in
Fig.~\ref{fig:dist_plots} the time
evolution of parameters for an individual plasma as it propagates to the center of the chamber before merging. Notably, $\beta$, $\omega_i \tau_i$, and $\omega_e \tau_e$ are anticipated to be equal to or greater than unity for an individual plasma during propagation and expansion into the chamber.  A key question is whether this is also true
upon merging, where densities can be much higher than for an individual plasma.

	\subsection{Negligible effect of neutrals}
	
	In this subsection, we show that the effect of neutrals should be negligible.
	
	First, we consider ion-neutral collisionality (dominated by charge exchange) within the plasmas before merging. The plasmas generated from the coaxial injectors are highly ionized. We utilize the PrismSpect atomic modeling software \citep{MacFarlane2004} with the temperature and density conditions in Table~\ref{tab:param} to estimate the fraction of neutral particles to be $<3 \times 10^{-6}$ near the gun and $<2 \times 10^{-6}$ near the chamber center. Within each plasma, the ion-neutral collision time is $\tau_{CX} = \nu_{CX}^{-1} \approx (n_{n} \sigma_{CX} v_{i,n})^{-1}$, where we assume that the relative speed between ions ($T_{i}=30$~eV) and cold neutrals is $v_{i,n}=54$~km/s. Based on the high ionization fraction, the neutral density is $n_{n} < 1.5 \times 10^{10}$~cm$^{-3}$. The H$^{1+}$-H$^{0}$ charge exchange cross section is $\sigma_{CX} \sim 3 \times 10^{-15}$~cm$^{2}$ \citep{Smirnov2000}. Therefore, we estimate that $\tau_{CX} \approx 4$~ms, much larger than the experimental duration. Comparatively, the thermal ion-ion collision times $\tau_i$ are $\sim$68~ns near the gun and $\sim$34~$\mu$s near the chamber center. Thus, ion-neutral collisionality within the plasma is a small effect.
	
	Next we consider the collisionality between the plasma ions and the background neutrals in the vacuum chamber. We expect plasmas injected into the chamber ($\sim 70$~km/s) to run ahead of injected neutrals (at room temperature 0.026~eV, thermal velocity $v_{th,n} < 1.5$~km/s). The vacuum chamber base pressure $\sim4~\mu$Torr corresponds to a neutral background density of $n_{n} \approx 1.3 \times 10^{11}$~cm$^{-3}$. The collision time for plasma ions colliding with the stationary background neutrals is $\tau_{background} = \nu_{background}^{-1} \approx (n_{n} \sigma_{CX} v_{i})^{-1}$, where $v_{i} \sim$70~km/s is the ion speed. We calculate this collision time to be $\tau_{background} \approx 400~\mu$s, again larger than the experimental duration. Comparatively, the ion-ion collision time for the two counter-streaming merging plasmas is approximately equal to the ion-ion slowing time $\tau_{ii,s} = \nu_{ii,s}^{-1} \approx (v_{rel}/4L_{ii,s})^{-1}$ \citep{Messer2013, Merritt2014}, where $v_{rel} = 2 v_{i} \sim$140~km/s is the relative plasma velocity and $L_{ii,s} \approx$ 57~cm is the ion-ion slowing length, which gives $\tau_{ii,s} \approx$ 16~$\mu$s. This calculation is a characteristic slowing time for H-H merging; for H-He and He-H the calculated slowing times are 3~$\mu$s and 27~$\mu$s, respectively. Thus, the effect of neutral collisionality is also negligible in the merging between opposing plasmas.

\subsection{Measurements near chamber center \label{sec:results_cent}}

\begin{figure}
	\includegraphics[width=\textwidth,keepaspectratio]{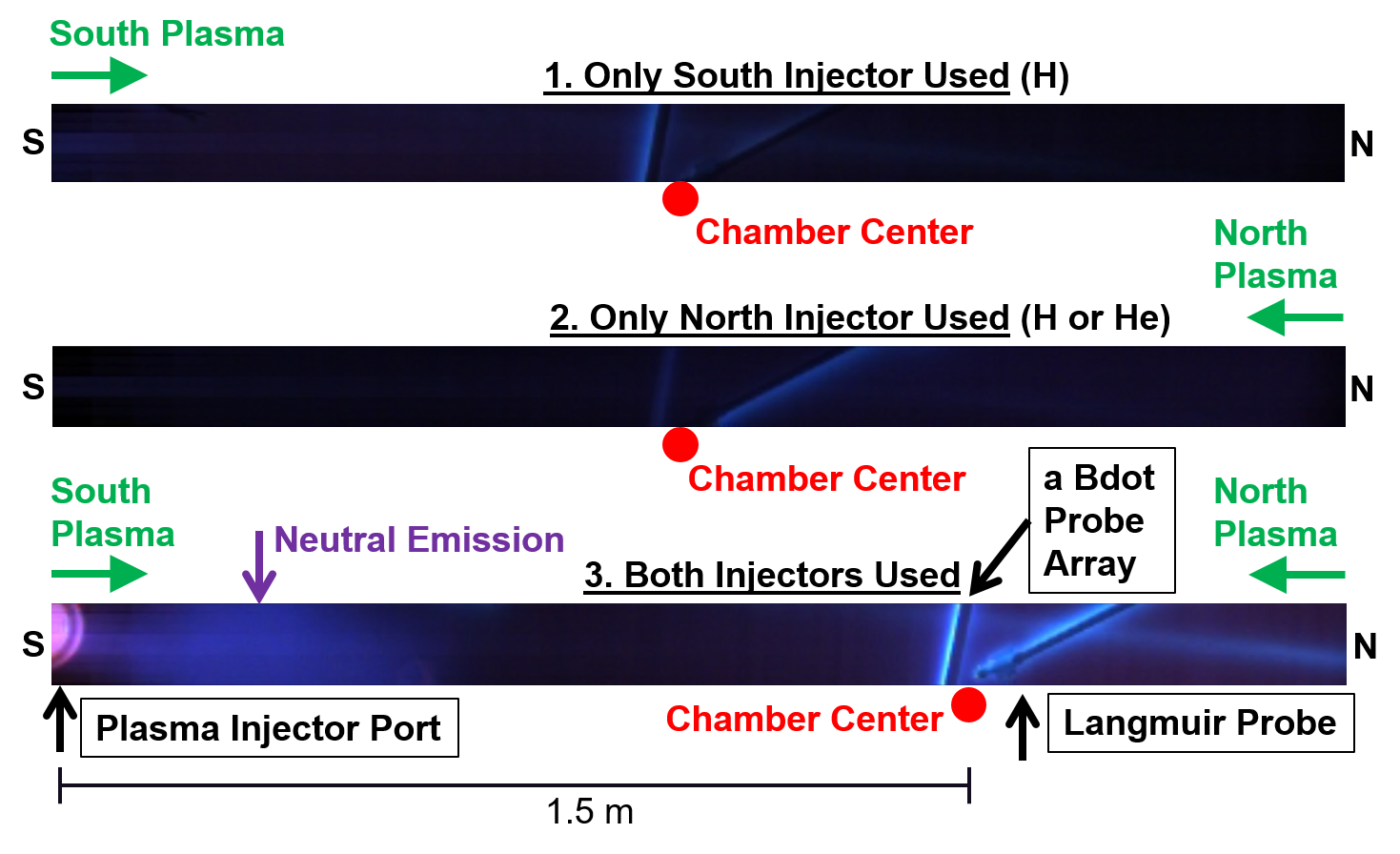}
	\caption{
		Fast-framing camera images of visible self-emission for plasmas from
		(top image)~South injector only, (middle image)~North injector only, and (bottom image)~both injectors. The main purpose of these images is to verify which sides of the probes light up due to the incoming plasma(s).
	}
	\label{fig:imaging}
\end{figure}

The primary results of this work are the measurement of plasma parameters--density, temperature, and magnetic field--from both individual plasmas and merged plasmas via head-on collisions, which occur near the ($r<$30~cm, $z=10$~cm) chamber position. We show evidence for obtaining transiently the desired state of $\beta>1$ and $\omega_{i}\tau_{i}>1$ for both situations. Figure~\ref{fig:imaging} shows visible imaging of individual plasmas (top two rows) coming from the South and North sides of the chamber, respectively, and also of a head-on plasma collision (bottom row) between the two individual plasmas. We use the emission seen on the sides of the probes to observe and verify from what direction the plasma is coming. Estimates of Bohm losses to probe surfaces are small compared to total particle inventories, 
and thus we expect the probes to not significantly perturb the plasma.

Here, the South H plasma has $\lambda_{gun,S} = 600$~m$^{-1}$, and the North He plasma has $\lambda_{gun,N} = 1200$~m$^{-1}$. Depending upon the formation and propagation dynamics, the ratio of toroidal to poloidal magnetic fields and corresponding  $\lambda_{plasma}$ for the plasma can be different from $\lambda_{gun}$, but $\lambda_{gun}$ influences the initial plasma state. Since $\lambda_{gun}$ satisfies $\infty > \lambda_{gun} > \lambda_{crit} = 92~$m$^{-1}$, we are in the desired parameter regime for obtaining $\beta>1$ and $\omega_{i}\tau_{i}>1$ plasmas. 

\begin{figure}
	\includegraphics[width=\textwidth,keepaspectratio]{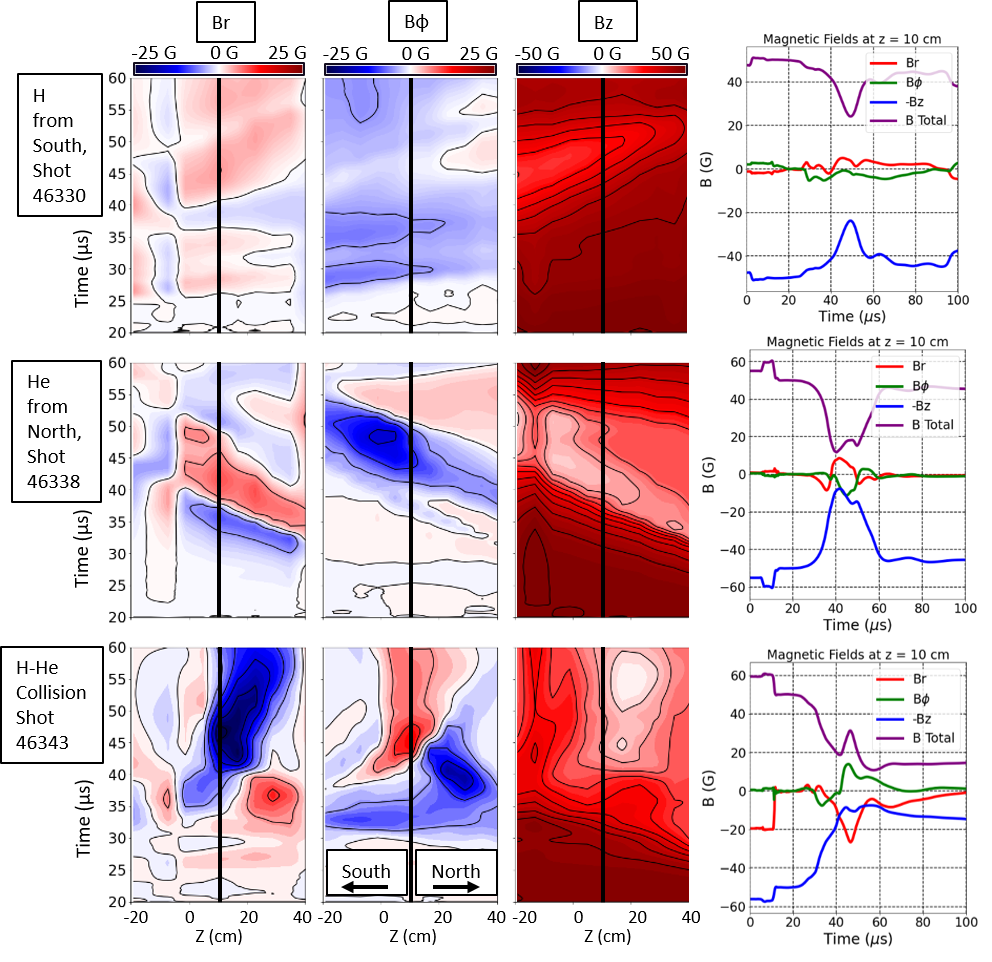}
	\caption{
		(Three left columns)~Magnetic-field components vs.\ $z$ (at $r=16$~cm)
		and $t$ for an individual H plasma (top row), an individual He plasma (middle row), and an H-He collision (bottom row). A signal traveling up and to the right is traveling through the chamber from South to North.  The contour lines represent 5-G increments. (Right column)
		Lineouts (corresponding to the vertical black lines in the contour plots)
		of magnetic-field components and total $B=(B_r^2 + B_\phi^2 + B_z^2)^{1/2}$ vs.\ $t$ at $z\approx 10$~cm (including the applied $B_z$ from the
		Helmholtz coil).  To avoid obscuring
		the $B$-total trace, $-B_z$ is plotted.
	}
	\label{fig:B_H_He}
\end{figure}

\begin{figure}
	\includegraphics[width=\textwidth,keepaspectratio]{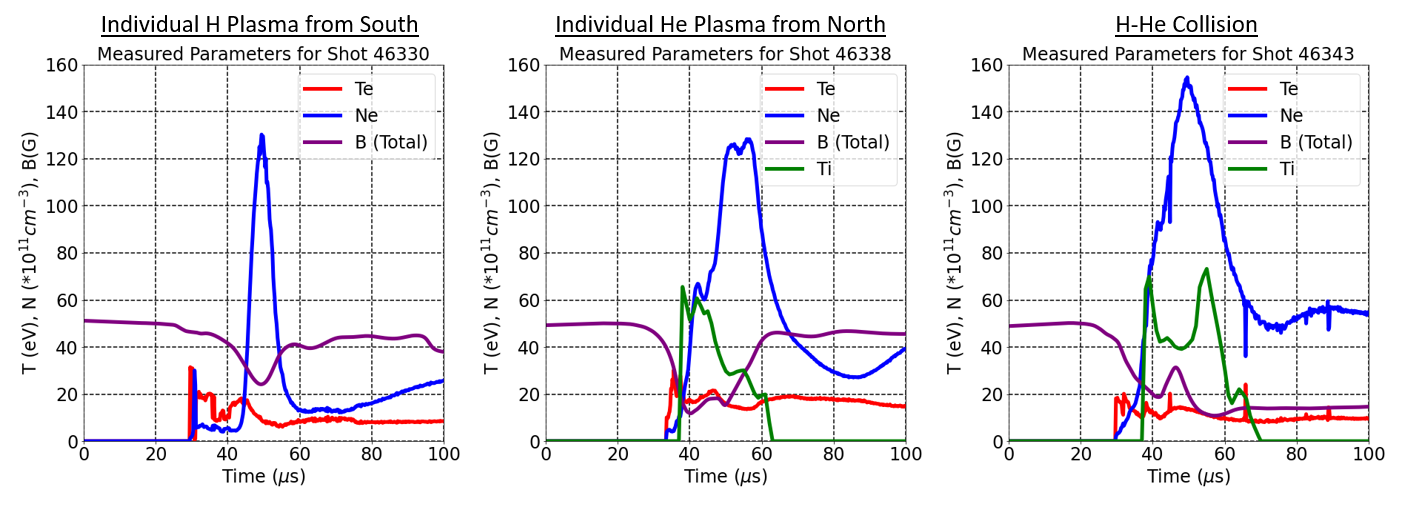}
	\caption{
		Plots of $T_e$ and $n_e$ (at $z=10$~cm, $r=25$~cm),
		total $B$ (at $z=10$~cm, $r=16$~cm, from Fig.~\ref{fig:B_H_He}), and
		$T_i$ (when He is present; radial viewing chord at $z=10$~cm) vs.\ $t$ for an (left) individual H
		plasma, (middle) individual He plasma, and (right) H-He collision.
	}
	\label{fig:param_H_He}
\end{figure}

Magnetic-field data are obtained by integrating the Bdot-probe-array signals. Figure~\ref{fig:B_H_He} shows the spatio-temporal evolution of the magnetic-field components for an individual H plasma from the South (shot 46330), an individual He plasma from the North (shot 46338), and for an H-He collision (shot 46343). 
From left to right, the three columns of colored contour plots are for the radial $B_r$, toroidal $B_\phi$, and axial $B_z$ fields ($z=0$ corresponds
to chamber center). For the contour plots, the Bdot-probe array is oriented
along $z$ and positioned at $r=16$~cm.  For reference, the arrival of the density signal can be seen in Fig.~\ref{fig:param_H_He}.

 Based on the slopes of the magnetic contours, we infer individual plasma axial velocities of $\sim 70$~km/s. As the plasmas propagate through the chamber and radially expand, they advect the 50-G background axial magnetic field in the radial direction, reducing the magnitude of the $B_{z}$ component within the plasma radius. In the H-He collision case, we observe larger magnetic-field strengths of the non-axial components ($\sim 30$~G compared to $<20$~G), and the axial field is reduced to $B_{z} <20$~G
 for a longer duration. Due to stagnation of the colliding plasmas, the magnetic signals last for $>40$~$\mu$s near chamber center for the H-He collision case
 compared to $\sim 20$~$\mu$s for the individual plasmas that propagate past
 chamber center. In future work, we intend to better characterize the anticipated diamagnetic effects based on the relative polarities of the background field compared with the magnetized plasmas.

Figure~\ref{fig:param_H_He} shows the time evolution of $T_e$ and $n_e$ (at $r=25~\mathrm{cm}, z= 10$~cm), and total $B$ (at $r=16~\mathrm{cm}, z= 10~\mathrm{cm}$, from Fig.~\ref{fig:B_H_He}) for an individual H plasma from the South (shot 46330), an individual He plasma from the North (shot 46338), and for an H-He collision (shot 46343). For shots 46338 and 46343, for which there is He present, 
$T_i$ is also plotted. Compared to the plasma from the North, the South plasma density persists for less time ($<20$~$\mu$s) and causes less reduction of the background magnetic field. The H-He collision case has a peak density above the value for the two individual plasmas, and at $>60$~$\mu$s, the magnetic signals decrease to $<15$~G, although density is still present. Additionally, $T_e$ for both the individual and colliding plasmas are $T_{e} \approx 15$~eV. The $T_i$ data show that $T_{i}>30$~eV$ > T_{e}$. At $>60$~$\mu$s, the $T_{i}$ signal falls to zero for low density and low electron temperature.

\begin{figure}
	\includegraphics[width=\textwidth,keepaspectratio]{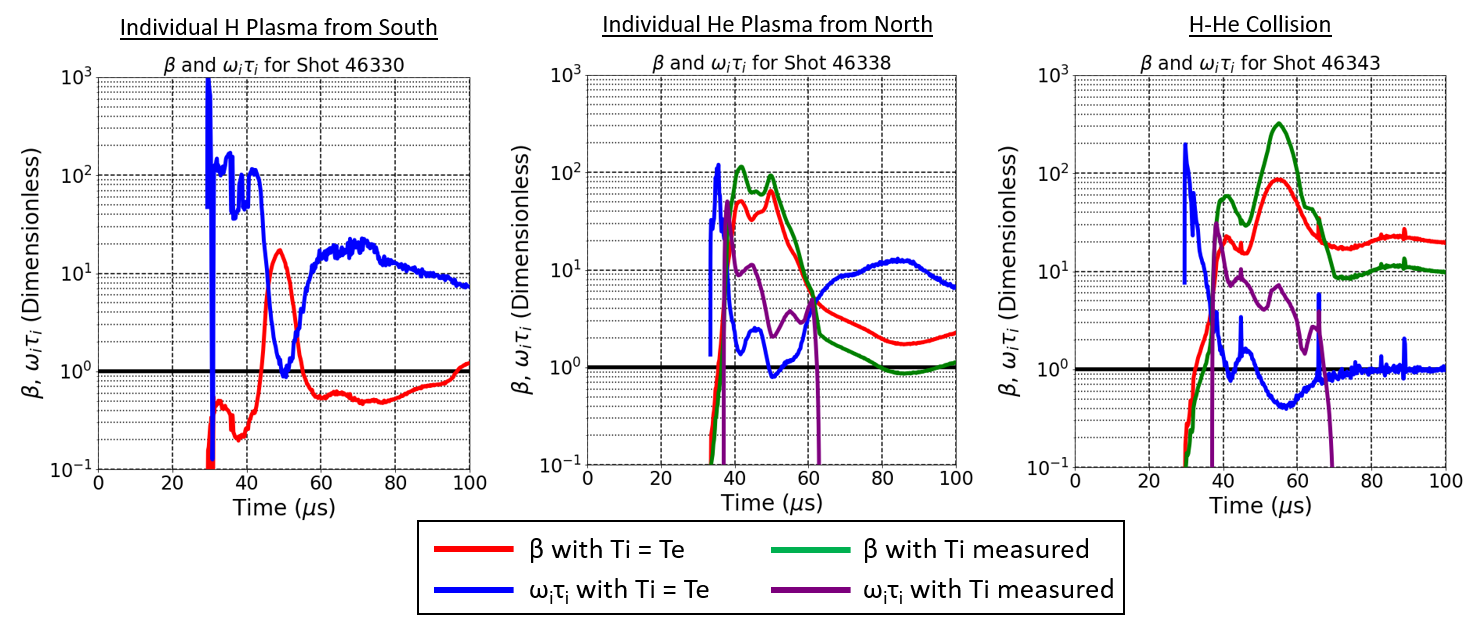}
	\caption{
		Derived values of $\beta$ and $\omega_i \tau_i$ vs.\ $t$, based on the quantities from Fig.~\ref{fig:param_H_He}, showing that there are time windows 
		up to 20~$\mu$s in duration during
		which $\beta$ and $\omega_i \tau_i$ are simultaneously greater than unity,
		for all three cases of (left) individual H plasma, (middle) individual He plasma, and (right) H-He collision.
	}
	\label{fig:beta_omegatau_H_He}
\end{figure}

Figure~\ref{fig:beta_omegatau_H_He} shows the derived values of $\beta$ and $\omega_{i}\tau_{i}$ vs.\ $t$, based on the quantities in Fig.~\ref{fig:param_H_He}. 
Because experiments with the individual plasma from the South (shot 46330) uses only H gas, we did not obtain direct $T_{i}$ measurements and therefore assume
$T_i=T_e$ for those shots (which serves as a lower bound for $T_{i}$, $\beta$, and $\omega_{i}\tau_{i}$). In the middle and right plots of Fig.~\ref{fig:beta_omegatau_H_He}, the red and blue traces (where $T_i=T_e$ is assumed) can be compared with the 
green and purple traces (based on measured $T_i$). We observe $\beta>1$ and $\omega_{i}\tau_{i}>1$ for $>20$~$\mu$s in both the individual H and He plasmas and in the H-He collision. In particular, at the 50-$\mu$s time of peak density in the collision, we obtain $\beta\approx 100$ and $\omega_{i}\tau_{i}\approx 4$ (using the measured $T_{i}$ values). Compared to the individual-plasma cases, which also obtain the desired $\beta,\omega\tau>1$ regime due to plasma expansion, the collision case gives
higher densities and durations which expands the parameter regime available for
study, and is more relevant for scaling to regimes of interest for an MIF target.

We find that the plasma from the North injector has larger $B_r$ and $B_\phi$ than the plasma from the South injector. The 50-G background $B_z$ from the Helmholtz coil is used to prevent radial expansion of the plasmas. Without the background field, we observe a significant decrease to near zero in the magnetic field signal in the plasma from the South injector, but we do still observe (smaller) magnetic signals in the plasma from the North injector. Studying the plasma collision without the background magnetic field is useful to support the future goals of studying collisions of more than two plasmas. In a possible future case of colliding six plasmas together (e.g., three orthogonal sets of head-on collisions), a Helmholtz-coil magnetic field would not be able to align with the propagation direction for six plasmas.

\section{Discussion and future work \label{sec:future}}

We intend to build on this work in upcoming experimental campaigns by (1)~exploring the time dynamics for the plasma collisions, especially looking into how the energy is distributed within the system, (2)~obtaining a spatial map of plasma parameters by scanning probe positions within the chamber, and (3)~performing a greater number of
shots per position to obtain statistics on density, temperature, and magnetic-field fluctuations.  During the experimental campaigns reported here, we did not complete a probe spatial scan because the individual plasmas did not form as reproducibly as desired. A likely reason for the unrepeatable nature is pre-ionization that is not toroidally symmetric within the coaxial injector. Future work will test more-uniform gas injection and pre-ionization systems. Preliminary engineering improvements using a new more-uniform annular pre-ionization system, a reduced gas injection timing delay from 3~ms to 2~ms, and four gas injectors instead of one around the annular gun nozzle have shown promise in creating more-reproducible plasmas with $\sim$95\% reliability for obtaining similar plasma parameters shot-to-shot. Equivalent levels of engineering improvements to plasma guns and their subsystems have recently been demonstrated in \citep{Yates2020}, which reported significantly improved balance of plasma mass injection in six- and seven-gun experiments as compared to \citep{Hsu2018}. Continuation of these experiments on the BRB will focus on better characterization of and generation of small-scale, tangled fields in these merged, $\beta>1,\omega\tau>1$ plasmas, as well as the exploration and variation of additional parameters, such as radiative cooling, magnetic Prandtl number, etc., of relevance to fundamental plasma physics and plasma-astrophysics questions.

Future studies of processes related to astrophysical phenomena including magnetohydrodynamic turbulence and magnetic dynamo will depend upon the plasma characteristics. Required resolution of spatial scales and time scales for MHD turbulence depend upon the Alfv\'en speed. From the approximate parameters obtained in the present work (see Table~\ref{tab:param}), a 50-cm-scale plasma environment should last $>35~\mu$s to support at least one crossing time associated with a $\sim14$~km/s Alfv\'en speed. The plasma needs to support $\sim$10 ion inertial lengths (0.3-7~cm) and ion gyroradii (0.3-30~cm) to allow turbulent cascades to develop. These turbulent cascades will be experimentally characterized while covering over two orders of magnitude in spatial scale by conducting millimeter-scale measurements. For studies of the magnetic dynamo, we plan to obtain large-enough ($>>1$) magnetic Prandtl numbers (the ratio of magnetic Reynolds number to Reynolds number, $Pr=R_{m}/Re \propto T^{4}$) and support at least one global turnover time over which the 50-cm-scale plasma can be advected by $\sim$50~km/s flows, in $\sim$10$\mu$s. 

Meanwhile, upcoming experiments on the Plasma Liner Experiment (PLX) \citep{Hsu2018, Yates2020} at Los Alamos National Laboratory will explore the merging of six or more plasmas to create $\beta>1,\omega\tau>1$ plasmas, at higher densities and temperatures than on the BRB, as a potential MIF target. To advance this MIF concept, future work on the BRB will explore formation of tangled magnetic fields within $\beta>1,\omega\tau>1$ plasmas, for which the magnetic field connection length is much longer than the local magnetic field scale size. These studies will evaluate thermal transport properties compared to theoretical predictions for diffusion coefficients in plasmas with tangled magnetic fields \citep{Kadomtsev1978,Rechester1978,chandran98}.

\section{Conclusions \label{sec:concl}}

In this work, we successfully used coaxial plasma-gun injectors to study
individual and head-on-merging plasmas on the BRB at the WiPPL.  By tuning the 
injector parameter $\lambda_{gun}
= \mu_0 I_{gun}/\psi_{gun}$
within an intermediate range between $\lambda_{crit} < \lambda_{gun} < \infty$, 
we were able to demonstrate the formation of a transient plasma with both $\beta>1$ and $\omega_{i}\tau_{i}>1$.  This is a promising first step toward establishing
a new experimental platform for studying frontier, weakly magnetized, high-$\beta$ plasma physics, e.g. the topics described in Sec.~\ref{sec:future}.  Compared with individual plasmas, the head-on collisions (1)~increase the duration for which the desired plasma state exists at a particular location and (2)~increase the magnitudes of $n$, $B_r$, and $B_\phi$, which widens the
parameter space compared to that achievable with individual plasmas.

\acknowledgments{T. Byvank and D. A. Endrizzi contributed equally to these experiments. The authors gratefully acknowledge J. Olson, J. Wallace, D. Den Hartog, and the rest of the WiPPL team for their technical support. We acknowledge H. Gota, T. Roche, I. Allfrey, and TAE Technologies for providing one of the
plasma injectors and its design. LANL staff were supported by the LANL Laboratory Directed Research and Development (LDRD) program under DOE contract no.\ 89233218CNA000001.  University of Wisconsin staff and the WiPPL user facility are supported by the DOE Office of Science, Fusion Energy Sciences under Award No.\ DE-SC0018266.  BRB construction was supported by a National Science Foundation Major Research Instrumentation grant.
 }

\bibliographystyle{jpp}
\bibliography{arxiv_BRBcol_2}

\end{document}